# Two-impurity Kondo effect in potassium doped single-layer *p*-sexiphenyl films

W. Chen[1], Y. J. Yan[2,3*], M. Q. Ren[1], T. Zhang[1,3,4], D. L. Feng[2,3,4]

[1]*State Key Laboratory of Surface Physics, Department of Physics, Laboratory of Advanced Materials, Fudan University, Shanghai, 200433, China*
[2]*Hefei National Laboratory for Physical Sciences at the Microscale and School of Physics, University of Science and Technology of China, Hefei, 230026, China*
[3]*Collaborative Innovation Center of Advanced Microstructures, Nanjing, 210093, China*
[4]*Shanghai Research Center for Quantum Sciences, Shanghai 201315, China*

Corresponding author：yanyj87@ustc.edu.cn

**ABSTRACT**

We show that a self-assembled phase of potassium (K) doped single-layer para-sexiphenyl (PSP) film on gold substrate is an excellent platform for studying the two-impurity Kondo model. On K-doped PSP molecules well separated from others, we find a Kondo resonance peak near $E_F$ with a Kondo temperature of about 30 K. The Kondo resonance peak splits when another K-doped PSP molecule is present in the vicinity, and the splitting gradually increases with the decreased inter-molecular distance, with no signs of phase transition. Our data demonstrate how a Kondo singlet state gradually evolves into an antiferromagnetic singlet state due to the competition between Kondo screening and antiferromagnetic RKKY coupling, as described in the two-impurity Kondo model. Intriguingly, the antiferromagnetic singlet is destroyed quickly upon increasing temperature and transforms back to a Kondo singlet well below the Kondo temperature. Our data provide a comprehensive picture and quantitative constraints on related theories and calculations of two-impurity Kondo model.

## I. INTRODUCTION

Kondo effect arises from the interactions between an isolated magnetic impurity and surrounding conduction electrons in a metal. The spin of the magnetic impurity is completely screened by conduction electrons, and a Kondo singlet state forms below the Kondo temperature [1]. It is commonly present in condensed-matter systems like individual magnetic atom [2-8] or molecule [9-15] on metals, and quantum dots [16-25]. On the other hand, when the magnetic atoms or ions form an ordered Kondo lattice, as in various heavy fermion crystals, magnetic sites could interact with each other by the RKKY interaction, which is mediated through the conduction electrons. The competition between Kondo screening effect and RKKY exchange interaction ($J$) in a Kondo lattice gives rise to a variety of complicated and interesting phenomena, such as ferromagnetic or antiferromagnetic orders, heavy fermions, and quantum criticality [26-28].

Both the single-impurity Kondo model and the Kondo lattice model have been explored extensively in theory and in experiment, but the intermediate regime between these two cases was addressed to a much less extent, particularly on the experimental side, despite its importance for the comprehensive understanding of Kondo physics. The two-impurity Kondo model (2IKM) is a typical case in this regime, where two Kondo impurities are coupled by RKKY interactions [23-25,29-37]. Theoretically, the ground state of the two-impurity Kondo system is determined by the relative strengths of the inter-impurity magnetic interaction and single-impurity Kondo temperature. It was predicted that for systems with particle-hole symmetry, a quantum phase transition will occur from the Kondo resonance state or Kondo singlet to the antiferromagnetic (AFM) singlet state, when inter-impurity $J$ is AFM and above a critical value $J^*$ [31-36]. In the absence of particle-hole symmetry, a crossover, rather than a quantum phase transition, is predicted between these two states [33-36]. On the other hand, if $J$ is ferromagnetic, another Kondo phase with higher spin may develop.

There are only a handful of experimental reports on 2IKM. In coupled double quantum dots (DQD) system [23-25] or two magnetic atoms/molecules on a metal surface [37], only spectra at several points in the parameter space are sampled. Controllable tuning of parameters is challenging, which also makes quantitative analysis difficult. Recently, a continuous tuning of the interatomic coupling was made possible in a scanning tunneling microscope (STM), by placing one Co atom on the STM tip and the other Co atom on a metal surface. By varying the tip–sample separation, the system can be smoothly driven from a Kondo singlet phase to an AFM phase [38]. However, the two magnetic impurities here are in different environments, not as in a usually 2IKM setup with RKKY interactions.

In this paper, we present a new platform to study the 2IKM, in a self-assembled ordered phase of potassium doped single-layer para-sexiphenyl (hereafter labeled as *p*-sexiphenyl or PSP) films grown on Au(111) substrate. We study the transformation between a Kondo singlet state to an AFM singlet state as a function of intermolecular distance (which controls the exchange interaction strength), and the temperature. Our results give a comprehensive phase diagram of the 2IKM, which deepens our understanding of this important many-body model system.

## II.  EXPERIMENTAL DETAILS

The experiments were performed in a low-temperature STM with ultrahigh vacuum and a base temperature of 5 K. Au(111) substrates were cleaned by cycles of Ar$^+$ sputtering and thermal annealing at 650 °C. PSP films of one monolayer (ML) thick were prepared by depositing PSP molecules (95% powder from Alfa Aesar) on the Au(111) surface kept at room temperature, using a Knudsen cell at 285 °C for about 4 minutes. The PSP source has been degassed thoroughly in advance to remove the impurity molecules (mainly the isomers of PSP molecule). K atoms were subsequently doped into the films using alkali metal dispensers, and the deposition rate of K was calibrated in advance by directly counting K atoms deposited on Au(111) under the same condition. The sample quality was further improved by annealing at room temperature for 30-60 minutes. STM topography was taken in a constant current mode, and the d$I$/d$V$ spectra were collected using a standard lock-in technique with a modulation frequency $f$ = 943 Hz. Pt-Ir tips were used for all the STM measurements after being treated on a clean Au(111) surface. All experimental data were collected at about 5 K, except for the temperature dependent experiments.

## III. RESULTS AND DISCUSSION

Figures 1(a,b) present STM images of a single-layer PSP film grown on the Au(111) substrate (labeled as PSP/Au). The PSP molecules self-assemble into a stripe-like structure, similar to other phenyl-based molecules adsorbed on metal substrates [39-45]. The stripe-like structure exhibits a periodicity of 2.95 ± 0.05 nm, while the inter-molecular distance along the stripe direction is 0.70 ± 0.05 nm. The pristine PSP film is insulating as discussed in Sec. 1 of the Supplementary Materials (SM) [46]. Subsequently, K atoms were deposited onto the PSP/Au films, and various phases with distinct structural and electronic properties were obtained with different K coverages which is quantified as $Kc$ in the unit of the areal density of PSP molecules in single-layer PSP/Au film (~ 5.14×10$^{13}$ cm$^{-2}$). One of these phases, labelled as phase I, is shown in Fig. 1(c) with $Kc$=3.39, which is the focus of this paper, while other ordered phases are discussed in Sec. 2 of SM [46].

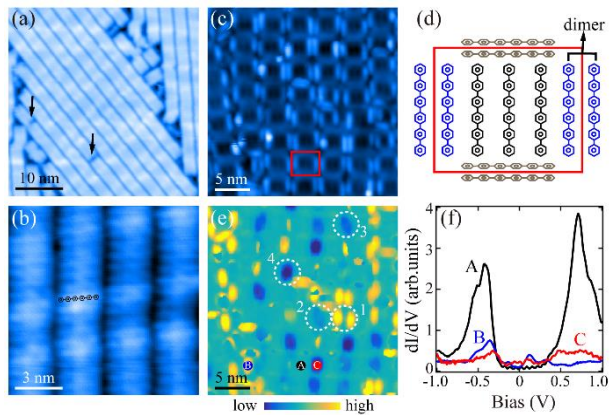

**FIG. 1. Structural and electronic properties of PSP/Au film and phase I in 3.39 ML K-doped PSP/Au film.** (a) A typical STM image of PSP/Au film ($V_b$=3.5 V, $I$=30 pA). The herringbone reconstruction (22 × √3) of Au (111) can still be distinguished on the film surface. Two molecular vacancies are marked by black arrows. (b) An enlarged STM image, in which single PSP molecule can be resolved ($V_b$=2.0 V, $I$=50 pA). (c) A typical STM image of phase I in 3.39 ML K-doped

PSP/Au film ($V_b$ = 3.0 V, $I$ = 30 pA). (d) Sketch of the possible lattice structure of phase I. The unit cell of phase I is indicated by the red rectangles in panels (c,d). (e) ZBC map measured in the area shown in panel (c) ($V_b$ = 10 mV, $I$ = 100 pA, $\Delta V$ = 1.0 mV). Four PSP molecular dimers with distinct electronic states are marked by white dashed circles. (f) Large-energy-scale d$I$/d$V$ spectra measured on different kinds of PSP molecules as marked in panel (e) ($V_b$ = 800 mV, $I$ = 100 pA, $\Delta V$ = 10 mV).

Figure 1(c) and Fig. S3 in SM display the detailed lattice structure of phase I. The unit cell of this phase is indicated by the red rectangle with a size of about 3.98 ($\pm$0.05) $\times$ 3.37 ($\pm$0.05) nm$^2$. Assuming the same areal density of PSP molecules in pristine and doped PSP/Au films, such a unit cell contains about seven PSP molecules. Considering the detailed surface morphologies and the size of a single PSP molecule, the possible structure of phase I is proposed as shown in Fig. 1(d) -- five PSP molecules lie in parallel on the Au substrate, while other two are oriented along perpendicular direction. About 24 potassium atoms (corresponding to the coverage $K_c$=3.39, i.e., 3.39 K atoms per PSP molecule) are intercalated into such a unit cell, but their exact distribution cannot be directly determined by STM.

Figure 1(e) displays a zero-bias conductance (ZBC) map taken in the same region in Fig. 1(c). For PSP molecules within a unit cell, most of them show featureless density of state (DOS) near $E_F$, except the two on the edges as marked out by blue color in Fig. 1(d) (see Sec. 4 of SM for details). For these two PSP molecules, we observe either a high DOS (bright yellow areas in Fig. 1(e)) or a low DOS (dark blue areas in Fig. 1(e)), depending on their molecular arrangements. Large-energy-scale d$I$/d$V$ spectra differ significantly on the molecules on the edge and in the center, as shown in Fig. 1(f). Such a difference in the electronic states suggests that the electron filling or the couplings between the molecules and the gold substrate vary strongly.

Figure 2 display four representative molecular dimers (referring to the two neighbor blue PSP molecules in Fig. 1(d)) with varying inter-molecular distance $d$ (as defined in Fig. 2(a)) and corresponding d$I$/d$V$ spectra. For molecular dimer #1, the two PSP molecules are far apart ($d$=2.34 nm), and sharp zero-bias peaks (ZBP) are observed on both of them, as shown in Fig. 2(a). The ZBP is the strongest right on the PSP molecules, and it gradually weakens as the tip moves away. However, when two PSP molecules are closer than those in dimer #1, such as dimers #2-#4 shown in Figs. 2(b-d), the ZBP splits into two peaks symmetrically with respect to $E_F$. The splitting magnitude strongly depends on the relative positions of the two edge molecules. For dimer #4 shown in Fig. 2(d), the two PSP molecules are very close and side by side, the splitting of the ZBP is the largest, reaching 8 ~ 10 meV. As for dimers #2 and #3, the two PSP molecules are staggered, the splitting becomes weaker.

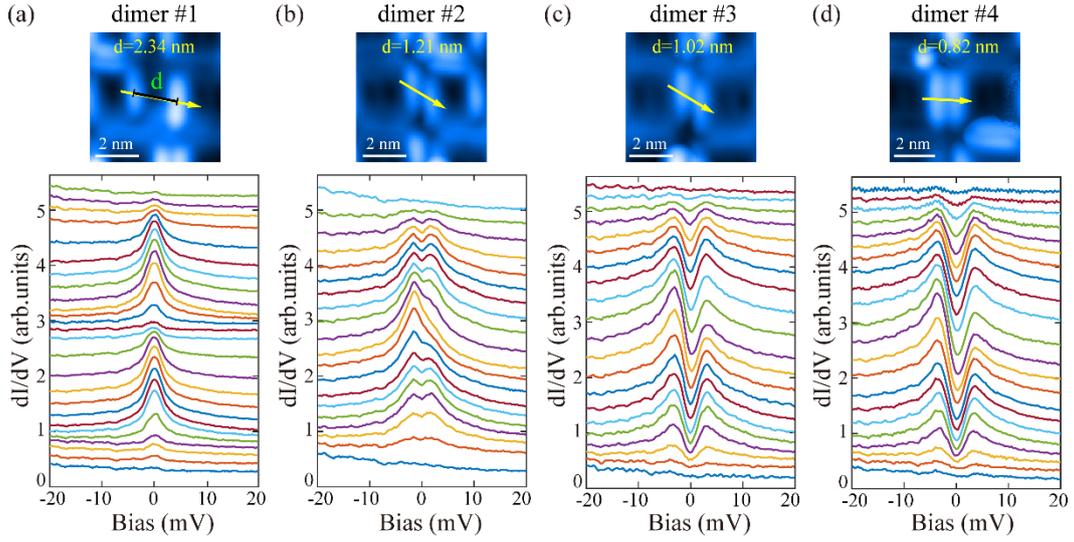

**FIG. 2. Distinct electronic states of PSP molecules in phase I.** Upper panels: STM images of four representative molecular dimers with inter-molecular distance $d$ labeled ($V_b$ = 0.2 V, $I$ = 30 pA); Lower panels: corresponding d$I$/d$V$ spectra collected along the yellow lines ($V_b$ = 10 mV, $I$ = 100 pA, $\Delta V$ = 1.0 mV). The inter-molecular distance, $d$, is defined as the spacing between centers of the two PSP molecules, as shown in the upper panel of (a).

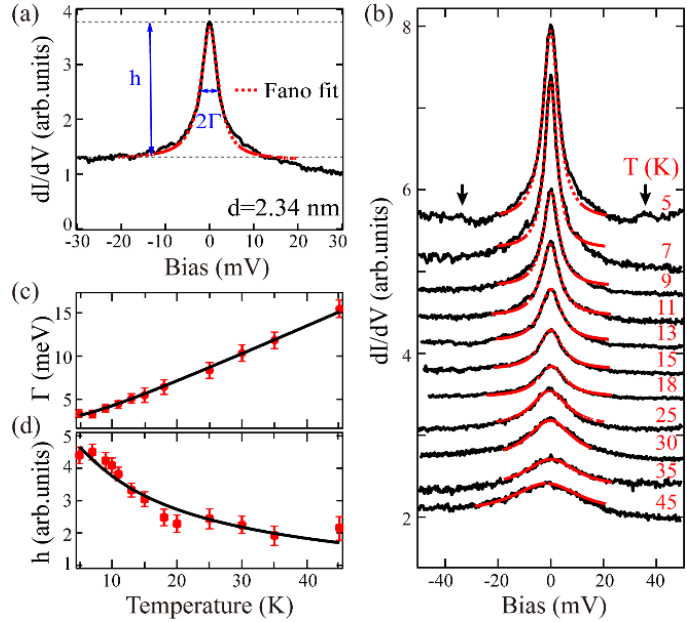

**FIG. 3. Properties of the ZBP.** (a) A typical ZBP spectrum that was well fitted by the Fano function, as indicated by the dashed red curve. The peak width and peak height are defined as $\Gamma$ and $h$, which are obtained through a Fano or Lorentzian fitting of the ZBP spectra. (b) Temperature dependence of the ZBP spectra. Red dashed curves are the fits to Fano functions. The black arrows indicate two symmetric side peaks arising from the inelastic-tunneling processes in which a molecular vibration mode is involved. (c,d) Temperature dependencies of $\Gamma$ and $h$, with the standard errors marked out. Fitting of the data is indicated by the black curves as described in the main text. Measurement conditions for the d$I$/d$V$ spectra here are: $V_b$ = 10 mV, $I$ = 100 pA, $\Delta V$ = 1.0 mV.

A ZBP observed on magnetic atoms or molecules on metal surfaces [2-15] and quantum dots [16-25] has been attributed to Kondo effect. Here we examine if the ZBP spectrum observed here fits in the Kondo singlet scenario as well. As shown in Fig. 3(a), the ZBP could be well fitted by a Fano lineshape, with a form factor $q$ of about 43. Such a large $q$ indicates that the tunneling into the likely Kondo resonance state dominates over tunneling directly into the metal substrate, facilitating an easy extraction of spectroscopic properties. As shown in Fig. 3(b), with increasing temperatures, the ZBP broadens and its intensity decreases quickly. The peak width, $\Gamma$, defined as the half-width at half-maximum (HWHM) of the ZBP, grows linearly with increasing temperature at higher temperature but tends to saturate at low temperature, as shown in Fig. 3(c). The linewidth broadening with temperature can be well fitted by an approximated expression developed in the framework of Fermi-liquid theory for a Kondo singlet: $\Gamma(T) = \sqrt{(\alpha k_B T)^2 + (2 k_B T_K)^2}$, where $k_B$ is Boltzmann's constant, $\alpha$ is an empirical parameter and $T_K$ is the Kondo temperature [8,14,47,48]. $\Gamma$ at zero-temperature defines the Kondo temperature $T_K=\Gamma/k_B$, thus $T_K$ here is determined as 30 K ±2 K for dimer #1. Moreover, temperature dependence of the conductance of a Kondo-induced peak can be fitted by an empirical expression that reproduces numerical renormalization group results for spin-1/2 systems:

$$h(T) = h_0 \left( \frac{T_K^2}{\left(2^{\frac{1}{s}} - 1\right) T^2 + T_K^2} \right)^s$$

, where $h_0$ is the $h$ value at zero temperature, $s$ is an empirical parameter of about 0.22 for spin-1/2 system [14,49]. As shown in Fig. 3(d), our data can be well fitted by this formula, which gives $T_K$ of about 30 K ± 3 K and $s$ of about 0.21 ± 0.02. The temperature dependence of the ZBP is thus consistent with the predicted behavior for a spin-1/2 Kondo effect. Based on these quantitative analyses, our data clearly indicate the observed ZBP can be attributed to the emergence of Kondo effect. The local moment here likely arises from an unpaired electron localized in an orbital of the PSP molecule due to K doping. In fact, the two step-like peaks at ±34 meV in the spectrum taken at 5 K in Fig. 3(b) have been identified as a phonon-assisted inelastic-tunneling feature of the Kondo effect in molecular systems, such as that of pure organic TTF-TCNQ molecules on the Au(111) surface [13] (see Sec. 5 of SM for further discussion).

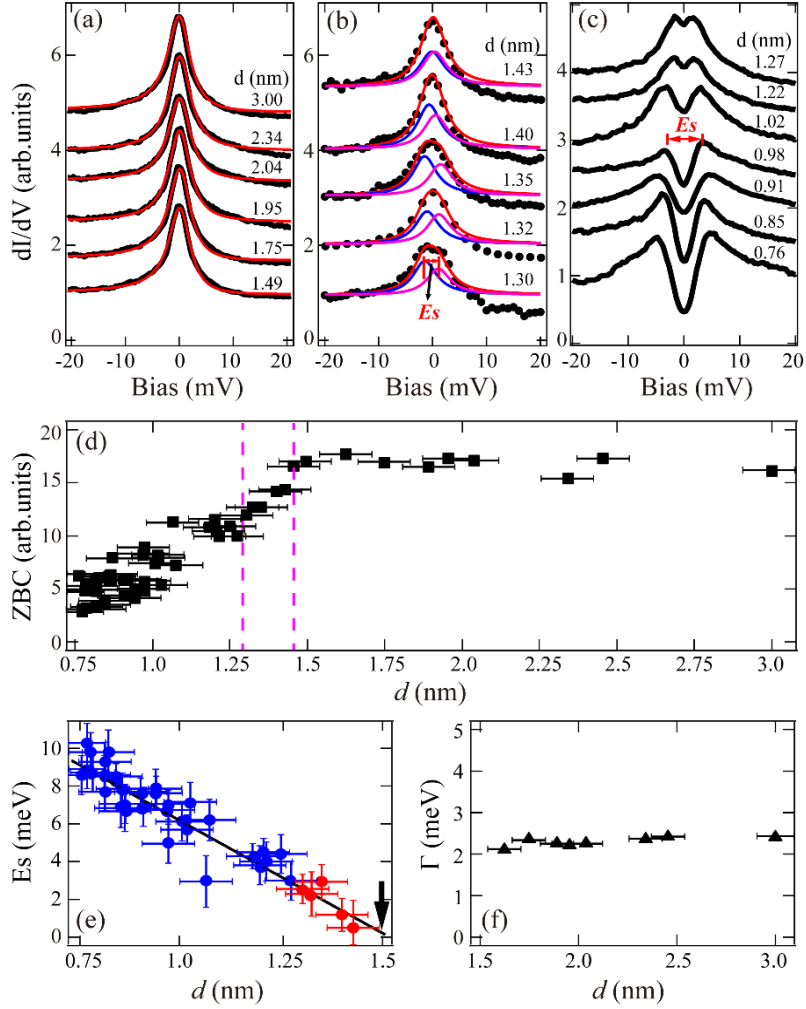

**FIG. 4. Inter-molecular distance dependence of spectral lineshape.** (a-c) Representative spectra collected on PSP dimers with decreasing inter-molecular distance, $d$, as labeled. The red double-arrowed line indicates the definition of the split magnitude (labelled as $E_S$), i.e., energy separation between the two split peaks. (d) ZBC as a function of $d$. (e) $E_S$ as a function of $d$, which can be well fitted by a linear function, and the black arrow indicates that the line extrapolates to zero at a critical value of $d \sim 1.5$ nm. Here, the data are obtained through double-Fano-function fitting of the spectra in panel (b) and directly read out the peak-separation of the spectra in panel (c). (f) HWHM of the Kondo resonance spectra ($\Gamma$) in panel (a), which are obtained by a Fano fitting of these spectra. Standard errors of the data are indicated by error bars.

The Kondo resonance of a K-doped PSP molecule makes it promising for studying how two Kondo singlets interact. As shown in Fig. 4(a), when the inter-molecular distance $d$ is above 1.49 nm, the Kondo peak remains intact with similar height and width (Figs. 4(d,f)). As $d$ decreases to below 1.49 nm, the peak starts to broaden with reduced height (Figs. 4(b,d)). Eventually, peak splitting can be clearly resolved when $d$ < 1.3 nm (Fig. 4(c)). The smallest $d$ we can observe is about 0.76 nm, likely limited by the finite size of PSP molecule. Fig. 4(d) plots the peak height (ZBC) as a function of $d$. The decreasing of the ZBC at $d$ < 1.49 nm indicates the peak starts to split. We therefore fitted the spectra of 1.43 nm < $d$ < 1.30 nm (Fig. 4(b)) with two peaks, whose width is fixed to a constant value similar to those in Fig. 4(f). The splitting magnitude ($E_S$, energy separation of the two split peaks or fitted double peaks) is shown in Fig. 4(e), which linearly decreases with increasing $d$. A linear fit of $E_S$ suggests that the splitting begins at about 1.5 nm.

In the previous studies of 2IKM, the competition between Kondo screening and antiferromagnetic RKKY interaction $J$ will cause the splitting of a Kondo resonance in the absence of magnetic field [29-38]. With increased antiferromagnetic $J/T_K$ the two single-impurity Kondo resonances gradually transform into an antiferromagnetic singlet state, manifesting as the splitting of ZBP. The splitting magnitude, $E_S$, is found to be $2J$ [29-38]. In our case, when two PSP molecules are far apart, they behave as Kondo impurities screened independently by itinerant electrons and show the usual Kondo resonance. When $d < 1.5$ nm, the inter-molecular interaction becomes obvious; the exchange interaction between localized molecule spins, likely RKKY antiferromagnetic coupling mediated by Au(111) substrate, cause the Kondo resonance peak to split. The well-formed double-peak spectrum at small $d$'s indicates the formation of an AFM singlet, although there is still some Kondo screening in the small $J$ regime, as described by the 2IKM.

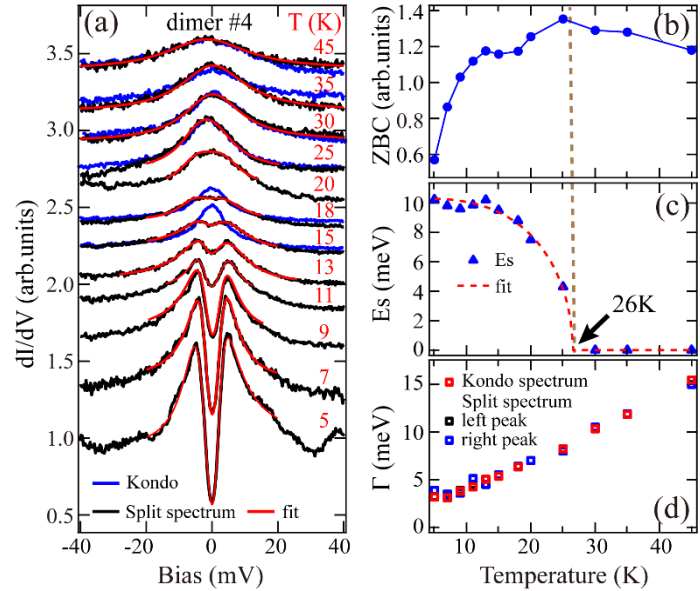

**FIG. 5. Temperature dependence of the double-peak spectrum.** (a) Temperature dependence of double-peak spectrum taken on dimer #4 (black curves, and measurement conditions for them are: $V_b = 10$ mV, $I = 100$ pA, $\varDelta V = 1.0$ mV), together with typical spectra taken at a Kondo singlet at the same labelled temperatures (blue curves). Red curves are the fits by double Fano-functions below 25 K and single Fano function above 25 K. (b,c) Temperature dependencies of ZBC and fitted $E_s$ of dimer #4. (d) Temperature dependence of the fitted $\varGamma$ for both dimer #4 and Kondo singlet data from a stand-alone molecule.

Figure 5(a) examines the temperature dependence of the AFM singlet state of dimer #4, whose $d = 0.82$ nm (Fig. 2(d)). As temperature increases, the splitting is suppressed quickly; the spectra at 18 K and 20 K still exhibit a double-peak lineshape, whereas the spectra taken at 25 K and above are almost identical to those Kondo peaks of a stand-alone molecule, which are overlaid on them. Meanwhile, the ZBC is a non-monotonic function of temperature, with a maximum at about 25 K (Fig. 5(b)). These signatures suggest that 25 K is the crossover temperature between a Kondo singlet and an AFM singlet. We have conducted a double-Fano-function fit to those data below 25 K (see Fig. S6 of SM), which gives the temperature dependence of $E_S$ in Fig. 5(c). The extrapolation in Fig. 5(c) consistently suggests $E_S$ diminishes at about 26 K. Therefore, our data indicate that the AFM singlet state evolves back to the Kondo singlet state above 26 K, and there is a transition from AFM to paramagnetic state above 26 K for these two local spins. This can be understood when considering that the Kondo screening exhibits a crossover behavior in temperature that persists well above $T_K$. In addition, an intriguing feature in Fig. 5(d) is that the peak width obtained from the double-Fano-function fit of the double-peak spectra and those from the fit to single Kondo resonance spectra in Fig. 3(c) are essentially the same and follow the same temperature dependence, which suggests the Kondo coupling is independent of the intermolecular distance. For dimer #2 with a smaller $E_S$ (Fig. 2(b)), similar temperature dependence is observed but the AFM singlet state transforms back to a Kondo singlet at a lower temperature of about 15 K (see Fig. S7 in SM).

RKKY coupling $J$ for a pair of molecular spins follows $J = J_0 \frac{\cos(2k_F d)}{d^3}$, where $J_0$ is a constant and $k_F$ is the Fermi wave vector of conductance electrons and $d$ is the distance between the molecules. Because the decreased $d$ monotonously corresponds to the increased $E_S$, or $2J$, the inter-molecular interaction should be antiferromagnetic in the studied range of 0.8-1.5 nm (Fig. 6(a)). We fit the experimental $J$'s to the above formula and obtained $k_F \sim 1.49$ nm$^{-1}$, close to the previous reported values of $k_F = 1.62$ nm$^{-1}$ and 1.85 nm$^{-1}$ considering the Rashba splitting of Au(111) surface states [50]. According to our fit, the $J$ above 1.5 nm is ferromagnetic and negligible small, it thus would not affect the Kondo singlet behavior for molecules parted further than 1.5 nm, consistent with the experiment.

The phase diagram presented in Fig. 6(b) summarizes our findings on this two-Kondo-impurity system. With increased AFM RKKY interactions, we observed a crossover from a Kondo singlet state to an AFM singlet -- the Kondo singlets start to interact when there is just minute amount of $J$, while the largest observed $J$ is about 5 meV or $1.95 k_B T_K$. We cannot determine if the AFM singlet state is partially formed (meaning there is still some Kondo screening as sketched in Fig. 6(b)) or fully developed with the largest achievable experimental $J$ here, which needs to be further studied in the future. Note that in theory, a quantum phase transition would occur at about $J^* \sim 2.2$-$2.5 k_B T_K$ between a Kondo singlet state and an AFM singlet state in systems with particle-hole symmetry [31-36]. Such a system has not been realized so far -- we did not observe a phase transition in the K-doped PSP molecules here either.

The temperature dependence is intriguing. As shown I Fig. 6(b), the transition temperatures between the AFM singlet (could be partially formed) at low temperatures and the Kondo singlet state at high temperatures for dimer #2 and #4 can be fit with a line going through the zero-point. This line defines a boundary between the Kondo singlet regime and the AFM singlet regime together with the crossover regime with partially formed Kondo singlet. At first sight, this is hard to rationalize within the context of the standard 2IKM [51]. As illustrated in Fig. 5, although the RKKY interaction $J$ is about $1.95 k_B T_K$ at 5 K for dimer #4, the AFM coupling of the two spins diminishes at just ~ $0.85 T_K$. However, it may be understood if considering a quickly diminishing RKKY interaction with increased temperature and the persistent nature of Kondo screening above $T_K$.

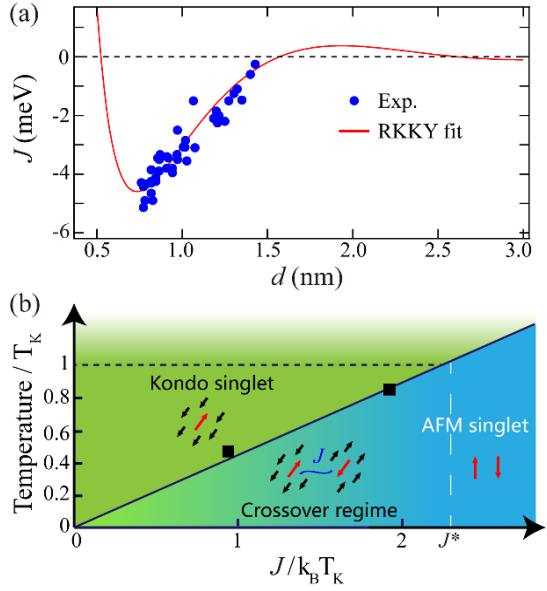

**FIG. 6. The RKKY exchange interaction and phase diagram of 2IKM found in K-doped PSP molecules.** (a) RKKY fitting of the experimental exchange interaction $J$ vs $d$. (b) Schematic phase diagram of our experimental 2IKM. The squares are the experimentally determined transition temperatures between the AFM singlet and the Kondo singlet for dimers #2 and #4 with increased temperature. The green region is the Kondo singlet region, and above $T_K$, the Kondo screening is significant weakened. The blue region is for well-developed AFM singlet. There is a broad crossover regime in the lower $J$ range, where the system evolves from a pure Kondo singlet into the AFM singlet. Here, a splitting spectral lineshape emerges with a small exchange interaction strength. The white dashed line is the theoretical predicated critical exchange interaction $J^* \sim 2.2\text{-}2.5 k_B T_K$, when there is particle-hole symmetry, which is not the case here.

## IV. CONCLUSION

We find that one of the self-assembled ordered phases of K-doped single-layer PSP/Au films is an excellent platform to study the 2IKM. One can study the transformation between a Kondo singlet and an AFM singlet as a function of temperature and RKKY interaction strength. Our findings depict a comprehensive experimental phase diagram of the 2IKM, highlighting the subtle interplay between RKKY interaction and Kondo screening. The revealed experimental energy and temperature scales put constraints on theories and calculations of the 2IKM. In addition, our findings suggest that more complex Kondo clusters and even a two-dimensional Kondo lattice may be constructed with such molecular systems.

## ACKNOWLEDGMENTS

We thank X. G. Gong, Y. L. Zhang, S. Kirchner, Z. Y. Lu, Y. F. Yang and Y. L. Wang for helpful discussions. This work is supported by the Science Challenge Project (Grant No. TZ2016004), National Natural Science Foundation of China (Grants No. 11774060), National Key R&D Program of the MOST of China (Grants No. 2017YFA0303004, and No. 2017YFA0303104), and Shanghai Education Development Foundation and Shanghai Municipal Education Commission (Chenguang Program). Shanghai Municipal Science and Technology Major Project (Grant No. 2019SHZDZX01).

Supplementary Materials for

"Two-impurity Kondo effect in potassium doped single-layer *p*-sexiphenyl films"

W. Chen[1], Y. J. Yan[2,3*], M. Q. Ren[1], T. Zhang[1,3,4], D. L. Feng[2,3,4]

[1]*State Key Laboratory of Surface Physics, Department of Physics, Laboratory of Advanced Materials, Fudan University, Shanghai, 200433, China*
[2]*Hefei National Laboratory for Physical Sciences at the Microscale and School of Physics, University of Science and Technology of China, Hefei, 230026, China*
[3]*Collaborative Innovation Center of Advanced Microstructures, Nanjing, 210093, China*
[4]*Shanghai Research Center for Quantum Sciences, Shanghai 201315, China*

## 1. Typical d*I*/d*V* spectra of the pristine PSP/Au films

Figure S1 presents the typical d*I*/d*V* spectra of 1 ML PSP/Au films within different energy scales, the data on bare Au(111) are also listed for comparison. There is a remarkable resemblance of the spectral lineshape between Au(111) and PSP/Au, suggesting that the PSP film is insulating and the low-energy DOS is mainly contributed by Au substrate.

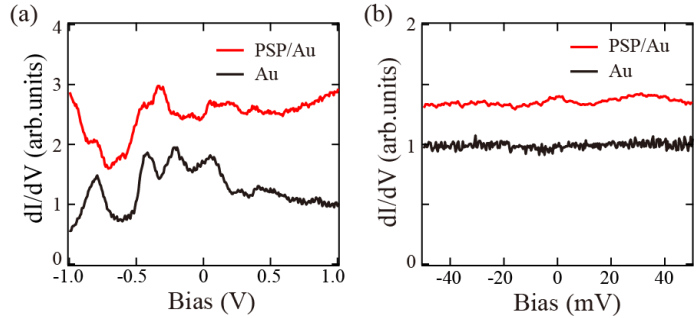

**FIG. S1. Typical d*I*/d*V* spectra of 1 ML PSP/Au films and bare Au (111) within different energy scales.** Curves in panels (a) and (b) are shifted vertically for clarity. Measurement condition: (a) $V_b$=0.2 V, *I*=50 pA, *ΔV*=30 mV, (b) $V_b$=10 mV, *I*=100 pA, *ΔV*=1 mV.

## 2. Abundant phases with various K coverages

When K atoms were doped into the PSP/Au film, abundant phases with distinct structural and electronic properties were obtained, as shown in Fig. S2. Similar to the *p*-terphenyl and *p*-quaterphenyl cases [1, 2], phase separation exists commonly due to the inhomogeneous distribution of K atoms. Fig. S2 shows the detailed STM morphology and low-bias d*I*/d*V* spectra of phases II-VII. For phases II-VI, the low-bias d*I*/d*V* spectra are featureless. While for phase VII, three kinds of low-bias d*I*/d*V* spectra, including a featureless DOS, a Kondo resonance peak and a weak double-peak feature as shown in Fig. S2(g), are observed on the corresponding regions shown in Fig. S2(f).

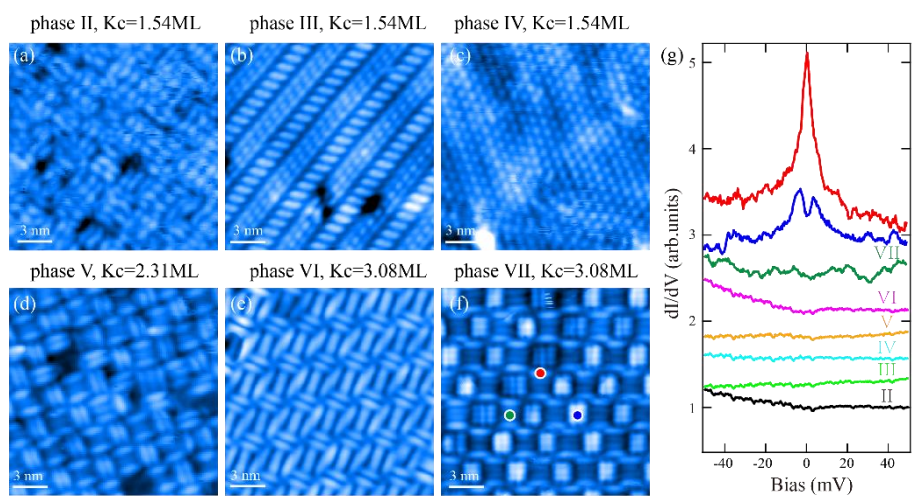

**FIG. S2. Structural and electronic properties of phases II–VII observed in PSP/Au films with various K coverages.** (a-f) STM images of phases II–VII. Measurement conditions: (a) $V_b = 2$ V, $I = 30$ pA; (b) $V_b = 2$ V, $I = 30$ pA; (c) $V_b = 1$ V, $I = 20$ pA; (d) $V_b = 0.5$ V, $I = 30$ pA; (e) $V_b = 0.2$ V, $I = 30$ pA; (f) $V_b = 0.2$ V, $I = 20$ pA. (g) Representative low-bias $dI/dV$ spectra of phases II–VII (Measurement condition: $V_b = 20$ mV, $I = 100$ pA, $\Delta V = 1$ mV). Curves are normalized to their values at +50 mV, and are shifted vertically for clarity. The circles with different colors in panel (f) indicate the positions where the top three spectra in panel (g) are measured.

## 3. STM images of phase I under different sample bias voltages

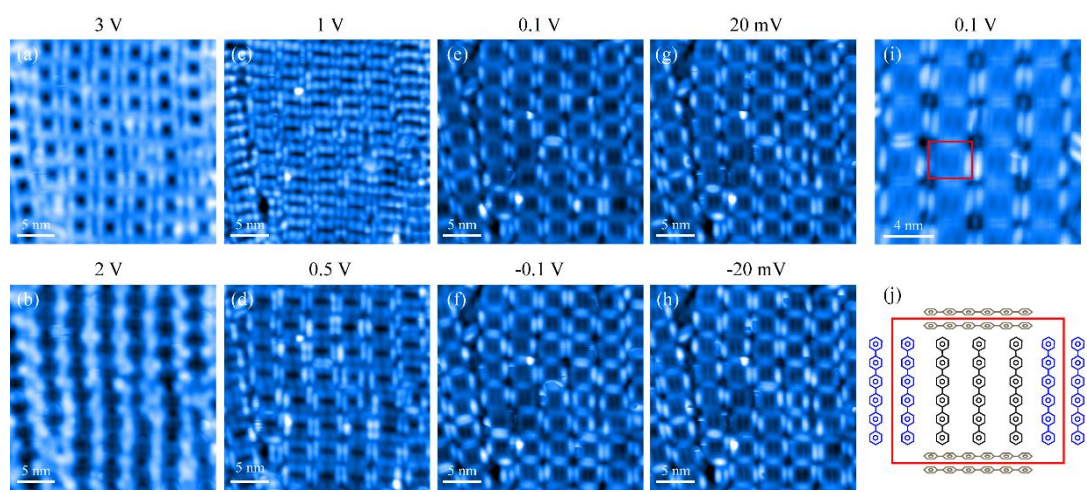

**FIG. S3. Structural properties of phase I.** (a-h) STM images of phase I measured under different bias voltages. (i) Detailed molecular pattern of phase I (Measurement condition: $V_b = 0.1$ V, $I = 50$ pA). (j) Sketch of the possible lattice structure of phase I. The unit cell of phase I is indicated by the red rectangles in panels (i) and (j).

## 4. Distinct electronic states of PSP molecules within one unit-cell of phase I

Figure S4 show the typical d$I$/d$V$ spectra collected on different PSP molecules within one unit-cell of phase I. Kondo spectra or split spectra can only be observed on the two edge molecules (blue ones in Fig. S3(j)), while the other molecules within the unit-cell exhibit featureless DOS near $E_F$.

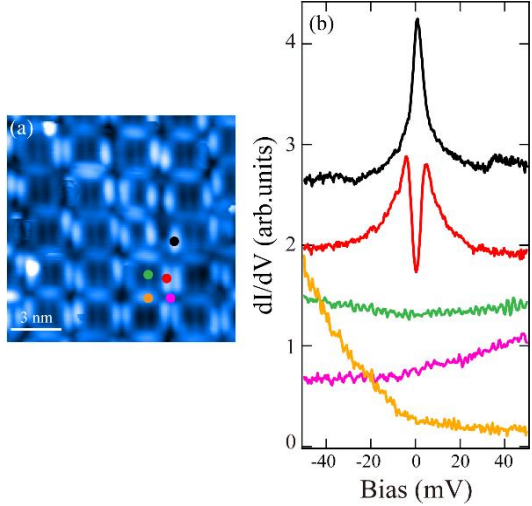

**FIG. S4. Typical dI/dV spectra for different PSP molecules within one unit-cell of phase I.** (a) A STM image of phase I. (b) dI/dV spectra measured on corresponding color-coded dots in panel (a) (Measurement condition: $V_b$ = 10 mV, $I$ = 100 pA, $\Delta V$ = 1.0 mV).

## 5. Discussion on the origin of the molecular spin

The origin of localized spin for the realization of Kondo effect in K-doped PSP molecules needs to be identified, since a single PSP molecule is non-magnetic and no magnetic atoms are introduced in our experiments. We notice a similar case in a single molecular layer of the TTF-TCNQ complex grown on an Au(111) surface, where Kondo effect is observed on the TCNQ molecules [3]. An unpaired electron, contributed by the TTF molecular donor, is localized in a π orbital of the TCNQ molecule. Due to the weak interaction between the TCNQ molecule and the underlying metal surface, a spin-1/2 degenerate ground state develops on the TCNQ molecule and results in the observation of Kondo effect. Because electrons in molecular orbitals strongly couple to molecular vibrations, a key feature for this unpaired-molecular-orbital-electron-induced Kondo effect is the two symmetric side peaks next to the Kondo peak and stepped increase of the d$I$/d$V$ signal that arise from phonon-assisted Kondo tunneling and inelastic tunneling processes [3-5]. Such a feature was observed in several molecular systems [3-7] and in our system as well. As shown in Fig. S5, symmetric side peaks and stepped d$I$/d$V$ increase at energies of about ±34~39 meV are observed on PSP molecules with Kondo singlet or AFM singlet state, while they weaken or disappear outside these molecules. The step-like features at ±34~39 meV can be attributed to the inelastic tunneling processes that absorb or excite a molecular vibration mode, as reported in ref. [3] for TCNQ and other molecular systems [6,7].

As K doping can give electrons to PSP molecules, we speculate that an unpaired electron contributed by K atoms may localize in a π orbital of the PSP molecule, and the weak interaction with the underlying metal surface, as suggested by the large Fano $q$, leads to the local moment. As shown in Fig. 1 and Fig. S2, the Kondo-related features are only observed in partial PSP molecules of phases I and VII, but they are absent in multiple other phases. As shown in Fig. 1(f), the electronic structures of molecules with and without zero-bias peak are very different. This suggests that a special molecular environment is required to realize the molecular local moment. The number of K atoms per molecule and the coupling strength between substrate and molecules, are considered important. For example, even number of electron filling or strong coupling between gold substrate and molecules may quench the molecular spin.

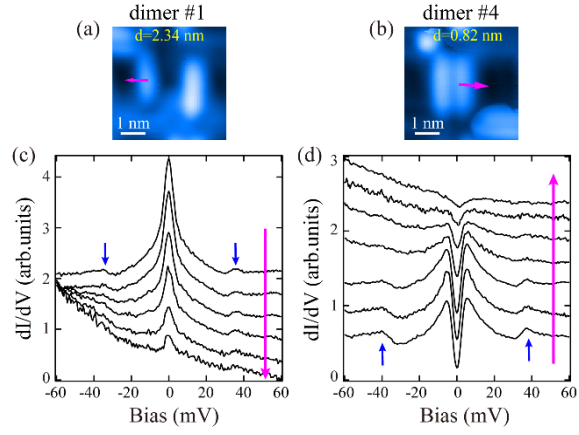

**FIG. S5. Manifestation of inelastic tunneling processes.** (a,b) STM images of molecular dimers #1 and #4. (c,d) Evolution of d$I$/d$V$ spectra within the energy range of ±60 mV measured along the two magenta linecuts in panels (a) and (b). The linecut directions are indicated by the arrowed magenta lines. Blue arrows point out the features of inelastic tunneling processes. Measurement conditions for the spectra are: $V_b$ = 10 mV, $I$ = 100 pA, $\Delta V$ = 2 mV.

6. **Fano fitting of the double-peak spectra measured under various temperatures**

The Kondo spectra in Fig. 3 and Fig. 4(a), the spectra in Fig. 5(a) and Fig. S6(b) above 25 K and the spectra in Fig. S7(a) above 15 K are fitted by a single Fano function:

$$f(x, q) = f_0 + \frac{(x+q)^2}{x^2+1},$$

$$x = \frac{E-E_K}{\Gamma},$$

with $E_K$ as the position and $\Gamma$ as the HWHM of the Kondo resonance peak, $q$ is the form factor that determines the spectral lineshape.

The spectra in Fig. 5(a) and Fig. S6 below 25 K and the spectra in Fig. S7(a) below 15 K are fitted by double Fano functions:

$$\rho(E) \propto a_1 f\left(\frac{E-E_L}{\Gamma_L}, q_1\right) + a_2 f\left(\frac{E-E_R}{\Gamma_R}, q_2\right).$$

Figs. S6 and S7 show the fitting results of the double-peak spectra measured on dimers #4 and #2 under various temperatures, respectively.

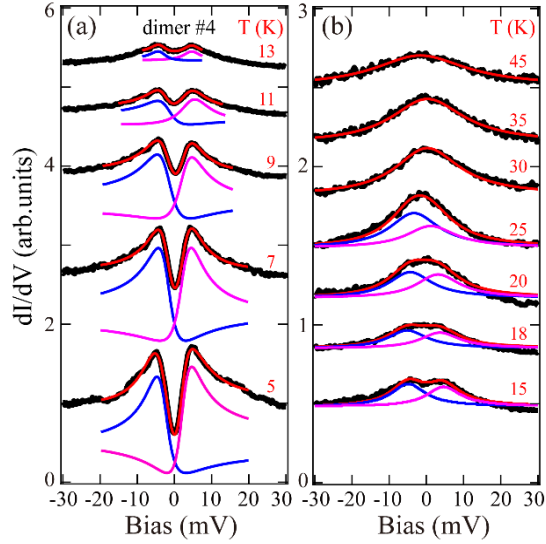

**FIG. S6. Fano fitting of the double-peak spectra measured on dimer #4 under various temperatures.** Panels (a) and (b) show the data at 5-13 K and 15-45 K, respectively. Red curves are the fits by double Fano functions below 25 K and by a single Fano function above 25 K.

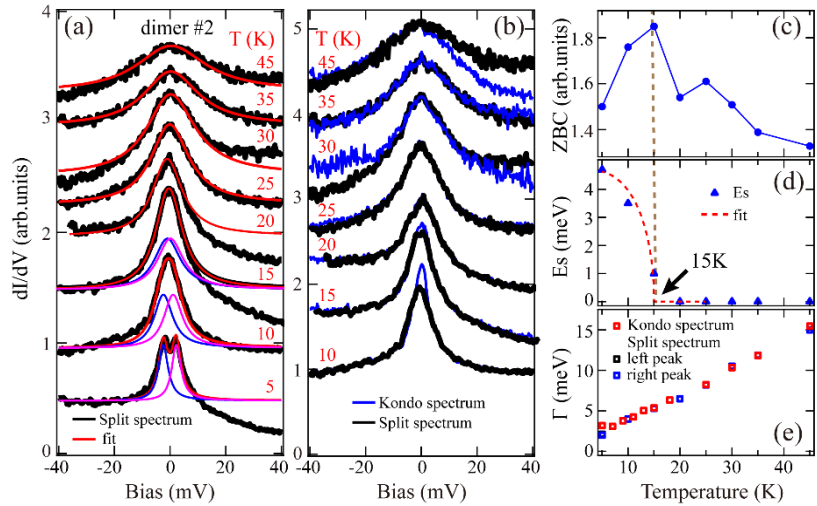

**FIG. S7. Temperature dependence of the double-peak spectrum measured on dimer #2.** (a) Temperature dependence of the double-peak spectra taken on dimer #2 with a smaller $E_S$ (black curves, and measurement conditions for them are: $V_b = 10$ mV, $I = 100$ pA, $\Delta V = 1.0$ mV). Red curves are the fits by double Fano-functions below 15 K and by a single Fano function above 15 K. (b) Comparison between the double-peak spectra taken on dimer #2 (black curves) and typical

spectra taken on a Kondo singlet at the same labelled temperatures (blue curves). (c,d) Temperature dependencies of ZBC and fitted $E_S$ of dimer #2. (e) Temperature dependence of fitted $\Gamma$ for both dimer #2 and Kondo singlet data from a stand-alone molecule.